\documentclass[a4paper]{article}
\RequirePackage[english]{babel}
\RequirePackage[latin1]{inputenc}
\RequirePackage[T1]{fontenc}
\RequirePackage{mathrsfs}
\RequirePackage{amsmath}
\RequirePackage{amssymb}
\RequirePackage{amsbsy}
\RequirePackage{bm}
\begin{document}
\title{\bf{Conformal Gravity\\ with Dirac Matter}}
\author{Luca Fabbri}
\date{}
\maketitle
\begin{abstract}
Recently, we have constructed the conformal gravity with metric and torsion, finding the gravitational field equations that give the conservation laws and trace condition; in the present paper we apply this theory to the case of the Dirac matter field: we shall see that in this case the trace condition becomes trivial, but further constraints arise, which will impose severe limitations on the dynamics of such a system.
\end{abstract}
\section*{Introduction}
That conformal gravity is important is due to a number of reasons: mathematically, its Lagrangian is unique as proven by Weyl, physically, its renormalizability was it was proven by Stelle \cite{s} and the fact it is ghost-free was recently demonstrated by Bender and Mannheim \cite{b-m}, and phenomenologically, it provides an explanation for dark matter as discussed by Mannheim and collaborators in a series of papers \cite{m-k,Mannheim:2010ti,Mann}; gravitational spontaneous conformal symmetry breaking has also been studied \cite{e-f-p/1,e-f-p/2}. But a complete theory of gravity possesses beside the curvature also the torsion tensor, arising as the gauge strengths of roto-translations in the Poincar\'{e} gauge theory \cite{h-h-k-n}; in this way the full coupling to both energy and spin density tensors may be established \cite{f}. Thus, metric as well as torsion conformal transformations must be defined \cite{sh}. In a recent paper, we have found a curvature with metric and torsion which is conformally invariant in the $(1+3)$-dimensional spacetime, finding the conformal metric-torsional theory by giving the system of gravitational field equations and the corresponding conservation laws and trace condition for spin and energy \cite{F,FABBRI}.

In this paper, we shall apply the above-mentioned theory of conformal gravity with curvature and torsion to the case of Dirac matter fields: after introducing the geometrical background we introduce the Dirac action; its variation will give the Dirac spin and energy densities with matter field equations, with which we will check the conservation laws. We will discuss the way in which the background reflects the constraints arising from the Dirac matter field itself.
\section{Conformal Curvature and Derivative}
In this paper we shall follow the notation of \cite{F}, and for matter as in \cite{FABBRI}.

In particular, the Riemann-Cartan metric-torsional geometry is defined in terms of a metric $g_{\alpha\beta}$ and a metric-compatible connection $\Gamma^{\mu}_{\alpha\sigma}$ such that metric and metric-compatible connection are independent: metric-compatibility means that by applying the covariant derivative associated to the connection onto the metric tensor the result vanishes; in general, the connection is not symmetric in the two lower indices, and so its antisymmetric part in those indices is a tensor that does not vanish, known as Cartan torsion tensor $Q_{\sigma\rho\alpha}$ decomposable as
\begin{eqnarray}
&\Gamma^{\sigma}_{\phantom{\sigma}\rho\alpha}=
\frac{1}{2}g^{\sigma\theta}[Q_{\rho\alpha\theta}+Q_{\alpha\rho\theta}+Q_{\theta\rho\alpha}
+(\partial_{\rho}g_{\alpha\theta}+\partial_{\alpha}g_{\rho\theta}-\partial_{\theta}g_{\rho\alpha})]
\label{connection}
\end{eqnarray}
showing that due to the presence of torsion metric and metric-compatible connections are independent. Next we introduce the Minkowskian metric $\eta_{ij}$ and a basis of vierbein $e_{\alpha}^{i}$ such that $e_{\alpha}^{p}e_{\nu}^{i}\eta_{pi}=g_{\alpha\nu}$ and a spin-connection $\omega^{ip}_{\phantom{ip}\alpha}$ such that vierbein and spin-connection are independent: metric-compatibilities spell that by applying the covariant derivative associated to the spin-connection to the Minkowskian metric and vierbein the results vanish; they respectively give the antisymmetry of the spin-connection $\omega^{ip}_{\phantom{ip}\alpha}=-\omega^{pi}_{\phantom{pi}\alpha}$ and the relationship
\begin{eqnarray}
&\omega^{i}_{\phantom{i}p\alpha}=
e^{i}_{\sigma}(\Gamma^{\sigma}_{\rho\alpha}e^{\rho}_{p}+\partial_{\alpha}e^{\sigma}_{p})
\label{spin-connection}
\end{eqnarray} 
showing again that vierbein and spin-connection are independent. The formalism with Greek letters and with Latin letters are respectively called spacetime and world formalism, and they are equivalent; in them, the independence between metric and connection is equivalent to the independence between vierbein and spin-connection. Although equivalent, these two formalisms differ for the fact that, while in spacetime formalism the transformations are the most general, in world formalism they are given in terms of Lorentz structure, which can be explicitly written in terms of other representations: the complex one is achieved by introducing the complex $\boldsymbol{\gamma}_{a}$ matrices verifying $\{\boldsymbol{\gamma}_{a},\boldsymbol{\gamma}_{b}\}=2\boldsymbol{\mathbb{I}}\eta_{ab}$ known as the Clifford algebra, from which one can define the complex $\boldsymbol{\sigma}_{ab}$ matrices defined to be given by $\boldsymbol{\sigma}_{ab}=\frac{1}{4}[\boldsymbol{\gamma}_{a},\boldsymbol{\gamma}_{b}]$ such that $\{\boldsymbol{\gamma}_{a},\boldsymbol{\sigma}_{bc}\}=i\varepsilon_{abcd} \boldsymbol{\gamma}\boldsymbol{\gamma}^{d}$ as the complex generators of the complex representation of the Lorentz algebra, called spinorial representation and needed to define matter fields: the spinor-connection 
\begin{eqnarray}
&\boldsymbol{\Omega}_{\rho}=\frac{1}{2}\omega^{ij}_{\phantom{ij}\rho}\boldsymbol{\sigma}_{ij}
\label{spinorial-connection}
\end{eqnarray} 
defines a spinorial covariant derivative with respect to which we have the constancy of the $\boldsymbol{\gamma}_{a}$ and therefore that of the matrices $\boldsymbol{\sigma}_{ab}$ automatically.

Before proceeding, we have to define another fundamental quantity given in terms of the connection or the spin-connection alone: Riemann curvature is
\begin{eqnarray}
\nonumber
&G^{i}_{\phantom{i}k\mu\nu}=G^{\rho}_{\phantom{\rho}\xi\mu\nu}e^{i}_{\rho}e^{\xi}_{k}
=(\partial_{\mu}\Gamma^{\rho}_{\xi\nu}-\partial_{\nu}\Gamma^{\rho}_{\xi\mu}
+\Gamma^{\rho}_{\sigma\mu}\Gamma^{\sigma}_{\xi\nu}
-\Gamma^{\rho}_{\sigma\nu}\Gamma^{\sigma}_{\xi\mu})e^{i}_{\rho}e^{\xi}_{k}\equiv\\
&\equiv\partial_{\mu}\omega^{i}_{\phantom{i}k\nu}-\partial_{\nu}\omega^{i}_{\phantom{i}k\mu}
+\omega^{i}_{\phantom{i}a\mu}\omega^{a}_{\phantom{a}k\nu}
-\omega^{i}_{\phantom{i}a\nu}\omega^{a}_{\phantom{a}k\mu}
\end{eqnarray}
and it is a tensor, antisymmetric in both the first and the second pair of indices, as it can be checked in terms of (\ref{connection}-\ref{spin-connection}). Also, we have that 
\begin{eqnarray}
\nonumber
&\boldsymbol{G}_{\mu\nu}=\frac{1}{2}G^{ij}_{\phantom{ij}\mu\nu}\boldsymbol{\sigma}_{ij}
=\frac{1}{2}(\partial_{\mu}\omega^{i}_{\phantom{i}k\nu}-\partial_{\nu}\omega^{i}_{\phantom{i}k\mu}
+\omega^{i}_{\phantom{i}a\mu}\omega^{a}_{\phantom{a}k\nu}
-\omega^{i}_{\phantom{i}a\nu}\omega^{a}_{\phantom{a}k\mu})\eta^{jk}\boldsymbol{\sigma}_{ij}\equiv\\
&\equiv\partial_{\mu}\boldsymbol{\Omega}_{\nu}-\partial_{\nu}\boldsymbol{\Omega}_{\mu}
+[\boldsymbol{\Omega}_{\mu},\boldsymbol{\Omega}_{\nu}]
\end{eqnarray} 
as it can be checked with (\ref{spinorial-connection}). For a complete introduction of the general geometric background without conformal structure we refer to \cite{f}.

The conformal transformations for all fields are given in terms of a unique function $\sigma$ and from the definition $\ln{\sigma}=\phi$, and the most general conformal transformation for torsion and the metric tensor is given by
\begin{eqnarray}
&Q^{\sigma}_{\phantom{\sigma}\rho\alpha}\rightarrow Q^{\sigma}_{\phantom{\sigma}\rho\alpha}
+q(\delta^{\sigma}_{\rho}\partial_{\alpha}\phi-\delta^{\sigma}_{\alpha}\partial_{\rho}\phi)\\
&g_{\alpha\beta}\rightarrow\sigma^{2}g_{\alpha\beta}
\end{eqnarray}
showing that the conformal transformation for torsion is actually a conformal transformation of its trace vector $Q_{\sigma}\!=\!Q^{\alpha}_{\phantom{\alpha}\alpha\sigma}$ alone $Q_{\alpha}\rightarrow Q_{\alpha}\!+\!3q\partial_{\alpha}\phi$ and then the conformal transformation for the metric-compatible connection will follow from relationship (\ref{connection}); the conformal transformation for the vierbein is thus 
\begin{eqnarray}
&e_{\alpha}^{k}\rightarrow\sigma e_{\alpha}^{k}
\end{eqnarray}
and conformal transformation for the spin-connection follows from (\ref{spin-connection}): finally because there is no conformal transformation for the constant $\boldsymbol{\gamma}_{a}$ matrices the conformal transformation for the spinor-connection is assigned automatically from (\ref{spinorial-connection}) while for the spinor $\psi$ and the dual spinor $\overline{\psi}\!=\!\psi^{\dagger}\boldsymbol{\gamma}_{0}$ it is given by
\begin{eqnarray}
&\psi\rightarrow\sigma^{-\frac{3}{2}}\psi\ \ \ \ \ \ \ \ \ \ \ \ \ \ \ \ \overline{\psi}\rightarrow\sigma^{-\frac{3}{2}}\overline{\psi}
\end{eqnarray}
which is the usual Dirac spinor field conformal transformation.

As it was already mentioned, we have discussed elsewhere that from the Riemann curvature with torsion it is not possible to extract its irreducible part finding it conformally invariant because of the presence of derivatives of torsion, but it is possible to introduce the modified metric-torsional curvature tensor as
\begin{eqnarray}
&M_{\alpha\beta\mu\nu}
=G_{\alpha\beta\mu\nu}+(\frac{1-q}{3q})(Q_{\beta}Q_{\alpha\mu\nu}-Q_{\alpha}Q_{\beta\mu\nu})
\label{curvature}
\end{eqnarray}
whose irreducible part given as usual by
\begin{eqnarray}
&T_{\alpha\beta\mu\nu}=M_{\alpha\beta\mu\nu}
-\frac{1}{2}(M_{\alpha[\mu}g_{\nu]\beta}-M_{\beta[\mu}g_{\nu]\alpha})
+\frac{1}{12}M(g_{\alpha[\mu}g_{\nu]\beta}-g_{\beta[\mu}g_{\nu]\alpha})
\label{conformalcurvature}
\end{eqnarray}
is conformally covariant in $(1+3)$-dimensional spacetimes. The commutator of spinorial covariant derivatives is given by the following usual expression
\begin{eqnarray}
&[\boldsymbol{D}_{\rho},\boldsymbol{D}_{\mu}]\psi
=Q^{\theta}_{\phantom{\theta}\rho\mu}\boldsymbol{D}_{\theta}\psi+\boldsymbol{G}_{\rho\mu}\psi
\label{conformalcommutator}
\end{eqnarray}
once the curvature is defined, and this is a geometric identity. It is worth mentioning that the conformal transformation for torsion is not uniquely defined, and as Shapiro discussed in \cite{sh}, there are two types of conformal transformations for torsion that can reasonably be defined: one is what he calls strong conformal transformation, given here by the transformation above, which is certainly a conformal transformation, since its effects are similar to those induced by the conformal transformation of the metric in the connection, as it can be witnessed from the fact that for the special value $q\!=\!1$ these contributions cancel each other exactly in such a way that the connection turns out to be conformally invariant; the other is what he calls weak conformal transformation, where torsion does not transform in any way whatsoever, and it is justified with the argument that torsion, being independent on the metric, does not have conformal transformations, which are a metric concept. The undetermined value of the constant $q$ is introduced for generality, and although the case $q\!=\!1$ is clearly special, it has to be notice that it is also trivial, since in this case the connection would be invariant, thus Riemann curvature would be invariant, there will be no need to define any irreducible curvature such as Weyl tensor and the theory will never recover the Weyl gravity in the torsionless limit; the value $q\!=\!0$ is also special, since that would make the strong conformal transformation reduce to the weak conformal transformation for torsion, but as (\ref{curvature}) shows here we cannot assume such a value, so that these two cases will have to be addressed in two independent situations. As we will not study a trivial case, and since the weak conformal case will be studied in a separate paper, here we will assume $q$ to have none of these values, and in fact no definite value. A general value of $q$ can be though as the fact that the conformal transformation for torsion and that induced by the conformal transformation of the metric in the connection are still structurally similar, although with a different weight: here we can think at $q$ as a conformal charge, much in the same way in which in the gauge theory of electrodynamics we think at the electric charge. With these comments we intend to clarify that such a strong conformal transformation for torsion is a reasonable option, though not the most reasonable one, and let alone the only one that is conceivable. But because this is nevertheless a reasonable choice, we will proceed to investigate some of its consequences in the following.
\section{Conformal Gravity and Matter: Dirac Fields}
Now that the background has been settled, we have to implement the dynamics, and we will follow again what we have done in reference \cite{F}.

To begin, we notice that as (\ref{conformalcurvature}) is a conformal tensor, then this is the tensor we have to use to determine the possible invariants: because of its symmetry properties and irreducibility, we have that the most general invariant is given according to the expression $AT^{\alpha\beta\mu\nu}T_{\alpha\beta\mu\nu}\!+\!BT^{\alpha\beta\mu\nu}T_{\mu\nu\alpha\beta}
\!+\!CT^{\alpha\beta\mu\nu}T_{\alpha\mu\beta\nu}$ in terms of the $A$, $B$, $C$ parameters, so that by defining the parametric quantity
\begin{eqnarray}
&P_{\alpha\beta\mu\nu}=AT_{\alpha\beta\mu\nu}+BT_{\mu\nu\alpha\beta}+\frac{C}{4}(T_{\alpha\mu\beta\nu}-T_{\beta\mu\alpha\nu}+T_{\beta\nu\alpha\mu}-T_{\alpha\nu\beta\mu})
\label{parametricconformalcurvature}
\end{eqnarray}
in terms of the parameters $A$, $B$, $C$, antisymmetric in the first and second pair of indices, irreducible and conformally covariant, such most general invariant reduces to the form given by $T^{\alpha\beta\mu\nu}P_{\alpha\beta\mu\nu}$ and so the most general action is
\begin{eqnarray}
&S=\int[kT^{\alpha\beta\mu\nu}P_{\alpha\beta\mu\nu}+L_{\mathrm{matter}}]\sqrt{|g|}dV
\end{eqnarray}
with $k$ gravitational constant, so far undetermined. By varying it we get
\begin{eqnarray}
\nonumber
&4k[D_{\rho}P^{\alpha\beta\mu\rho}+Q_{\rho}P^{\alpha\beta\mu\rho}
-\frac{1}{2}Q^{\mu}_{\phantom{\mu}\rho\theta}P^{\alpha\beta\rho\theta}-\\
&-(\frac{1-q}{3q})(Q_{\rho}P^{\rho[\alpha\beta]\mu}
-\frac{1}{2}Q_{\sigma\rho\theta}g^{\mu[\alpha}P^{\beta]\sigma\rho\theta})]=S^{\mu\alpha\beta}
\label{Fabbri}\\
\nonumber
&2k[P^{\theta\sigma\rho\alpha}T_{\theta\sigma\rho}^{\phantom{\theta\sigma\rho}\mu}
-\frac{1}{4}g^{\alpha\mu}P^{\theta\sigma\rho\beta}T_{\theta\sigma\rho\beta}
+P^{\mu\sigma\alpha\rho}M_{\sigma\rho}+\\
\nonumber
&+(\frac{1-q}{3q})
(D_{\nu}(2P^{\mu\rho\alpha\nu}Q_{\rho}
-g^{\mu\alpha}P^{\nu\theta\rho\sigma}Q_{\theta\rho\sigma}
+g^{\mu\nu}P^{\alpha\theta\rho\sigma}Q_{\theta\rho\sigma})+\\
&+Q_{\nu}(2P^{\mu\rho\alpha\nu}Q_{\rho}
-g^{\mu\alpha}P^{\nu\theta\rho\sigma}Q_{\theta\rho\sigma}
-P^{\mu\nu\rho\sigma}Q^{\alpha}_{\phantom{\alpha}\rho\sigma}))]=\frac{1}{2}T^{\alpha\mu}
\label{Weyl}
\end{eqnarray}
where $T^{\mu\nu}$ and $S^{\rho\mu\nu}$ are the energy and spin density tensors of the matter conformal field in general; if on the one hand Einstein equations describe how energy is the source of curvature while Sciama-Kibble equations describe how spin is the source of torsion, here both Weyl equations and this new set of equations describe how energy and spin are the source of an intertwined combination of both curvature and torsion. This fact will be dramatic for Dirac matter.

Finally, field equations (\ref{Fabbri}-\ref{Weyl}) imply the conservation laws given by
\begin{eqnarray}
&D_{\rho}S^{\rho\mu\nu}+Q_{\rho}S^{\rho\mu\nu}
+\frac{1}{2}T^{[\mu\nu]}=0
\label{conservationlawspin}\\
&D_{\mu}T^{\mu\rho}+Q_{\mu}T^{\mu\rho}-T_{\mu\sigma}Q^{\sigma\mu\rho}
+S_{\beta\mu\sigma}G^{\sigma\mu\beta\rho}=0
\label{conservationlawenergy}
\end{eqnarray}
with trace condition as another conservation law
\begin{eqnarray}
&(1-q)(D_{\mu}S_{\nu}^{\phantom{\nu}\nu\mu}+Q_{\mu}S_{\nu}^{\phantom{\nu}\nu\mu})
+\frac{1}{2}T_{\mu}^{\phantom{\mu}\mu}=0
\label{conservationlawtrace}
\end{eqnarray}
and these conservation laws will be satisfied once the matter conformal field equations are assigned: the energy is not traceless because its trace is related to the spin trace vector through (\ref{conservationlawtrace}); because the constraint constituted by the trace condition (\ref{conservationlawtrace}) is a conservation law then this constraint is dynamically implemented in the model. Again, in the following we will see that these dynamical properties of the trace condition will collapse for Dirac matter fields.

So to introduce matter fields, our choice will be that of picking the simplest but also the most important of all matter fields that are known, that is the massless Dirac field, which is already conformally invariant, with Dirac action
\begin{eqnarray}
&S=\int[L_{\mathrm{gravity}}
+\frac{i}{2}(\bar{\psi}\boldsymbol{\gamma}^{\rho}\boldsymbol{D}_{\rho}\psi
-\boldsymbol{D}_{\rho}\bar{\psi}\boldsymbol{\gamma}^{\rho}\psi)]|e|dV
\label{action}
\end{eqnarray}
where it is over the volume of spacetime that the integral is taken. By varying (\ref{action}) we get the completely antisymmetric irreducible spin and traceless energy densities given by the usual expressions as in the following
\begin{eqnarray}
&S_{\mu\alpha\beta}=
\frac{1}{4}\varepsilon_{\mu\alpha\beta\rho}
\bar{\psi}\boldsymbol{\gamma}^{\rho}\boldsymbol{\gamma}\psi
\label{spin}\\
&T_{\mu\alpha}=\frac{i}{2}(\bar{\psi}\boldsymbol{\gamma}_{\mu}\boldsymbol{D}_{\alpha}\psi
-\boldsymbol{D}_{\alpha}\bar{\psi}\boldsymbol{\gamma}_{\mu}\psi)
\label{energy}
\end{eqnarray}
along with the massless matter field equations
\begin{eqnarray}
&i\boldsymbol{\gamma}^{\mu}\boldsymbol{D}_{\mu}\psi
+\frac{i}{2}Q_{\mu}\boldsymbol{\gamma}^{\mu}\psi=0
\label{matterequations}
\end{eqnarray}
as a simple calculation would show, and as it is widely known.

As it is now possible to see, conservation laws (\ref{conservationlawspin}-\ref{conservationlawenergy}) and (\ref{conservationlawtrace}) are in fact satisfied by the spin and energy densities (\ref{spin}-\ref{energy}) so soon as the conformal matter field equations (\ref{matterequations}) are accounted; when a geometrical background is conformally invariant there is the loss of one degree of freedom realized by the introduction of the trace condition as constraint: for the Dirac field the complete antisymmetry of its spin implies its tracelessness, so that as we already mentioned the conservation law for the trace reduces to be non-dynamical any longer, which might have been regarded as a problem if it were not for the fact that its masslessness implied also the energy to be traceless, so that such constraint became trivial, and no problem is faced. Since Dirac fields in massless configuration are already conformally invariant, that the conservation law for the trace provides no additional information is to be expected after all.

However, things will not be so straightforward for the effects of the complete antisymmetry of the spin on the structure of the field equations, and to see that we write the entire set of field equations, which then reads according to
\begin{eqnarray}
\nonumber
&4k[D_{\rho}P^{\alpha\beta\mu\rho}+Q_{\rho}P^{\alpha\beta\mu\rho}
-\frac{1}{2}Q^{\mu}_{\phantom{\mu}\rho\theta}P^{\alpha\beta\rho\theta}
-(\frac{1-q}{3q})(Q_{\rho}P^{\rho[\alpha\beta]\mu})]=\\
&=\frac{1}{4}\varepsilon^{\mu\alpha\beta\rho}
\bar{\psi}\boldsymbol{\gamma}_{\rho}\boldsymbol{\gamma}\psi
\label{Fabbri-spin}\\
\nonumber
&2k[P^{\theta\sigma\rho\alpha}T_{\theta\sigma\rho}^{\phantom{\theta\sigma\rho}\mu}
-\frac{1}{4}g^{\alpha\mu}P^{\theta\sigma\rho\beta}T_{\theta\sigma\rho\beta}
+P^{\mu\sigma\alpha\rho}M_{\sigma\rho}+\\
\nonumber
&+(\frac{1-q}{3q})
(D_{\nu}(2P^{\mu\rho\alpha\nu}Q_{\rho})+Q_{\nu}(2P^{\mu\rho\alpha\nu}Q_{\rho}
-P^{\mu\nu\rho\sigma}Q^{\alpha}_{\phantom{\alpha}\rho\sigma}))]=\\
&=\frac{i}{4}(\bar{\psi}\boldsymbol{\gamma}^{\alpha}\boldsymbol{D}^{\mu}\psi
-\boldsymbol{D}^{\mu}\bar{\psi}\boldsymbol{\gamma}^{\alpha}\psi)
\label{Weyl-energy}
\end{eqnarray}
with the massless matter field equations
\begin{eqnarray}
&i\boldsymbol{\gamma}^{\mu}\boldsymbol{D}_{\mu}\psi
+\frac{i}{2}Q_{\mu}\boldsymbol{\gamma}^{\mu}\psi=0
\label{Dirac}
\end{eqnarray}
after some simplification has been performed as a consequence of the validity of the condition $P^{\mu\nu\rho\sigma}Q_{\nu\rho\sigma}=0$ coming from the irreducibility of the spin density tensor and therefore of its field equation: however the irreducibility of the spin density tensor is only a part of the more stringent constraint coming from having a completely antisymmetric spin density tensor and consequently only the completely antisymmetric form of the field equations must be accounted, hence implying that all other decompositions of the spin field equations must give rise to strong restrictions. Nevertheless, as we have already mentioned, there is a difference between the usual situation we had in Einstein gravitation and in this Weyl gravitation: in Einstein gravity, the field equation for the spin coupling to torsion was such that the complete antisymmetry of the spin was immediately translated into the complete antisymmetry of torsion; here in Weyl gravity, both field equations for the spin and energy couple to both torsion and curvature in such a way that the complete antisymmetry of the spin is partly imposed on torsion and partly on the curvature, so that there no longer is a completely antisymmetric torsion and additionally there are additional constrictions on the curvature tensor itself. What this will imply is that in the limit in which there is no torsion, while the Einstein-type of gravity would reduce to the Einstein gravity, the Weyl-type of gravity would reduce to the Weyl gravity complemented by subsidiary conditions for the curvature tensor, which are not necessarily verified in general instances, unless its solutions are restricted correspondingly.

To see that this is indeed the case, let us consider the fact that we may decompose torsional contributions away from the torsionless terms in all curvatures and derivatives: once this will be done, all curvatures and derivatives will remain written in terms of the purely metric curvature and derivatives given by the Weyl conformal tensor $C_{\mu\alpha\sigma\rho}$ and the Levi-Civita derivative $\boldsymbol{\nabla}_{\mu}$ plus contributions due to torsion decomposable in its three decompositions according to the form $Q_{\mu\alpha\sigma}\!\equiv\! T_{\mu\alpha\sigma}\!+\!\varepsilon_{\mu\alpha\sigma\rho}W^{\rho}\!+\!\frac{1}{3}(g_{\mu\alpha}Q_{\sigma}
\!-\!g_{\mu\sigma}Q_{\alpha})$ where $T_{\mu\alpha\sigma}$ is the non-completely antisymmetric irreducible part and $W^{\alpha}$ the axial vector dual of the completely antisymmetric irreducible part of torsion; because the conformal transformation law of torsion is entirely inherited by its trace then its two irreducible parts are conformally covariant, which is an important fact to know because in a conformal theory it is possible to require the vanishing of conformally covariant tensors alone. In general, the Dirac equation is
\begin{eqnarray}
&i\boldsymbol{\gamma}^{\mu}\boldsymbol{\nabla}_{\mu}\psi
-\frac{3}{4}W_{\mu}\boldsymbol{\gamma}^{\mu}\boldsymbol{\gamma}\psi=0
\end{eqnarray}
that is the form we would have had in Einstein-type of gravity although in Weyl-type of gravity there no longer is the possibility to substitute torsion with the spin of spinors and there no longer are non-linear self-interactions for spinors in the spinorial field equations: in this theory even in presence of torsion we have the linearity of the Dirac equation. The torsional completion of the metric conformal gravity for Dirac fields radically changes the whole dynamics.
\section*{Conclusion}
In this paper, we have considered the conformal theory of gravity with Dirac matter fields: we have discussed the fact that due to the complete antisymmetry of the spin and the masslessness of the Dirac field the conservation laws for the trace is trivial but some of the constraints coming from the complete antisymmetry of the spin do not yield the complete antisymmetry of torsion but they are transferred into constraints for the metric; we have also seen that the non-linear self-interactions of spinorial matter fields are absent. The fact that in conformal gravity the conformally invariant gravitational field equations have a very different structure compared to the non-conformal case implies that even if the spin and energy tensor and the matter field equations for the Dirac field are the same we would have had in the non-conformal case, the overall conformal dynamics of gravity with matter is radically different: even if torsion were removed, there would remain strong constraints over the metric, as well as over the structure of the matter field. This is quite an interesting result, because it means that when the full coupling to both torsion and metric is accounted in conformal gravity even an apparently natural conformal field such as the Dirac spinorial massless field is subject to severe limitations.

\end{document}